\documentclass[11pt]{article}

\usepackage{fullpage}

\newif\iflong
\longtrue
\long\def\comment#1{}

\newcommand{\dotprod}[2]{ \langle\! {#1} , {#2} \! \rangle}  
\renewcommand{\Pr}[1]{{\mathop{\null \mbox{\boldmath $P$}}\left[ #1 \right]}} 
\newcommand{\softo}{\tilde{O}}
\newcommand{\remove}[1]{}

\newcommand{\E}[1]{{\mathop{\null \mbox{\boldmath $E$}}\left[ #1 \right]}}  

\newlength{\figtxtwid}
\long\gdef\boxitnew#1{\dimen200 = \hsize \advance\dimen200 by -7pt
\begingroup\vbox{\hrule \hbox to \hsize{\vrule\kern3pt
      \vbox{\hsize \dimen200 \kern3pt#1\kern3pt}\hfil \kern3pt\vrule}\hrule}\endgroup}

\long\gdef\boxit#1{\begingroup\vbox{\hrule\hbox{\vrule\kern3pt
      \vbox{\kern3pt#1\kern3pt}\kern3pt\vrule}\hrule}\endgroup}

\newlength{\myfigwidth}
\newtheorem{theorem}{Theorem}[section]
\newtheorem{lemma}[theorem]{Lemma}
\newtheorem{corollary}[theorem]{Corollary}

\newtheorem{fact}[theorem]{Fact}
\newtheorem{definition}{Definition}[section]

\newcommand{\qed}{\mbox{}\hspace*{\fill}\nolinebreak\mbox{$\rule{0.7em}{0.7em}$}}
\newenvironment{proof}{\par{\bf Proof:}}{\(\qed\) \smallskip \par}{}

\newcommand{\marnote}[1]{\marginpar{\raggedright \footnotesize \em #1}}
\def\Olog{\tilde{O}}

\begin{document}

\begin{titlepage}
\title{\Large Approximate Graph Coloring by Semidefinite Programming
}
\author{{\sc David Karger} \thanks{MIT Laboratory for Computer Science,
    Cambridge, MA   02139.  Email {\tt karger@mit.edu.}  URL {\tt
      http://theory.lcs.mit.edu\~karger.} 
Supported by a Hertz Foundation Graduate
Fellowship; by NSF Young Investigator Award CCR-9357849, with
matching funds from IBM, Schlumberger Foundation, Shell Foundation and
Xerox Corporation; and by NSF Career Award CCR-9624239}
\and
{\sc Rajeev Motwani}
\thanks{Department of Computer Science, Stanford University, Stanford,
CA 94305 ({\tt rajeev@cs.stanford.edu}).
Supported by an Alfred P.~Sloan Research Fellowship, an IBM Faculty
Development Award, and NSF Young Investigator Award CCR-9357849, with
matching funds from IBM, Mitsubishi, Schlumberger Foundation, Shell
Foundation, and Xerox Corporation.}
\and
{\sc Madhu Sudan}
\thanks{MIT Laboratory for Computer Science,
    Cambridge, MA   02139.  Email {\tt madhu@lcs.mit.edu.}  
Work done when this author was at IBM's Thomas J. Watson Research 
Center, Yorktown Heights, NY 10598.
}
}
\date{}
\thispagestyle{empty}
\maketitle

\begin{abstract}
We consider the problem of coloring $k$-colorable graphs with the
fewest possible colors. We present a randomized polynomial time
algorithm that colors a 3-colorable graph on $n$ vertices with
$\min\{O(\Delta^{1/3}\log^{1/2}\Delta \log n),\allowbreak
 O(n^{1/4}\log^{1/2} n)\}$ 
colors
where $\Delta$ is the maximum degree of any vertex.  Besides giving
the best known approximation ratio in terms of $n$, this marks the
first non-trivial approximation result as a function of the maximum
degree $\Delta$.  This result can be generalized to $k$-colorable
graphs to obtain a coloring using 
$\min\{{O}(\Delta^{1-2/k}\log^{1/2}\Delta \log n),\allowbreak
{O}(n^{1-3/(k+1)}\log^{1/2} n)\}$ colors.  Our results are inspired by the
recent work of Goemans and Williamson who used an
algorithm for {\em semidefinite optimization problems}, which
generalize linear programs, to obtain
improved approximations for the MAX CUT and MAX 2-SAT problems.  An
intriguing outcome of our work is a duality relationship established
between the value of the optimum solution to our semidefinite program
and the Lov\'asz $\vartheta$-function.  We show lower bounds on the
gap between the optimum solution of our semidefinite program and the
actual chromatic number; by duality this also demonstrates interesting
new facts about the $\vartheta$-function.
\end{abstract}

\end{titlepage}

\section{Introduction}
\label{sec:intro}

A legal vertex coloring
of a graph $G(V,E)$ is an assignment of colors to its vertices such that
no two adjacent vertices receive the same color.  Equivalently, a
legal coloring of $G$ by $k$ colors is a partition of its vertices
into $k$ independent sets.  The minimum number of colors needed for
such a coloring is called the chromatic number of $G$, and is usually
denoted by $\chi(G)$.  Determining the chromatic number of a graph is
known to be NP-hard (cf.~\cite{GarJon}).

Besides its theoretical significance as a canonical NP-hard problem,
graph coloring arises naturally in a variety of applications such as
register allocation~\cite{briggs,chaitin,chandra} and
timetable/examination scheduling~\cite{berge,wood}. In many
applications that can be formulated as graph coloring problems, it
suffices to find an {\em approximately optimum} graph coloring---a
coloring of the graph with a small though non-optimum number of
colors.  This along with the apparent impossibility of an exact
solution has led to some interest in the problem of approximate graph
coloring.
The analysis of approximation algorithms for graph coloring
started with the work of Johnson \cite{jo:co} who shows that a version of
the greedy algorithm gives an $O(n/\log n)$-approximation algorithm
for $k$-coloring.
Wigderson~\cite{wigderson} improved this bound
by giving an elegant algorithm that uses
$O(n^{1-1/(k-1)})$ colors to legally color a $k$-colorable graph.
Subsequently, other polynomial time algorithms were provided by
Blum~\cite{blum94} that use $O(n^{3/8}\log^{8/5} n)$ colors to
legally color an $n$-vertex 3-colorable graph.  This result
generalizes to coloring a $k$-colorable graph with $O(n^{1-1/(k -
4/3)} \log^{8/5} n)$ colors.  The best known performance guarantee for
general graphs is due to Halld\'orsson~\cite{halldorsson} who provided a
polynomial time algorithm using a number of colors that is within a
factor of $O( n (\log \log n)^2/ \log^3 n)$ of the optimum.

Recent results in the hardness of approximations indicate that it may
be not possible to substantially improve the results described above.
Lund and Yannakakis~\cite{LuYa} used the results of Arora, Lund,
Motwani, Sudan, and Szegedy~\cite{almss} and Feige, Goldwasser,
Lov\'asz, Safra, and Szegedy~\cite{FGLSS} to show that there exists a
(small) constant $\epsilon > 0$ such that no polynomial time algorithm
can approximate the chromatic number of a graph to within a ratio of
$n^{\epsilon}$ unless P $=$ NP. The current hardness result for the
approximation of the chromatic number is due to 
Feige and Kilian \cite{FK} and 
H{\aa}stad \cite{hastad},
who show that approximating it to within 
$n^{1 - \delta}$, for any $\delta > 0$, would imply NP=RP (RP is the
class of probabilistic polynomial time algorithms making one-sided
error). 
However, none of these hardness
results apply to the special case of the problem where the input graph
is guaranteed to be $k$-colorable for some small $k$. The best
hardness result in this direction is due to Khanna, Linial, and
Safra~\cite{sanjeev} who show that it is not possible to color a
3-colorable graph with 4 colors in polynomial time unless P $=$ NP.

In this paper we present improvements on the result of Blum.  In
particular, we provide a randomized polynomial time algorithm that
colors a 3-colorable graph of maximum degree $\Delta$ with
$\min\{O(\Delta^{1/3} \log^{1/2}\Delta \log n),O(n^{1/4} \log^{1/2} n)\}$ 
colors; moreover, this
can be generalized to $k$-colorable graphs to obtain a coloring using
$O(\Delta^{1-2/k}\log^{1/2}\Delta \log n)$ or 
${O}(n^{1-3/(k+1)}\log^{1/2} n)$ colors.
Besides giving the best known
approximations in terms of $n$, our results are the first non-trivial
approximations given in terms of $\Delta$.
Our results are based on the recent work of Goemans and
Williamson~\cite{GW:maxcut} who used an algorithm for {\em
semidefinite optimization problems} (cf.~\cite{GLS,Alizadeh}) to
obtain improved approximations for the MAX CUT and MAX 2-SAT problems.
We follow their basic paradigm of using algorithms for semidefinite
programming to obtain an optimum solution to a relaxed version of the
problem, and a randomized strategy for ``rounding'' this solution to a
feasible but approximate solution to the original problem. Motwani and
Naor~\cite{cutcover} have shown that the approximate graph coloring
problem is closely related to the problem of finding a CUT COVER of
the edges of a graph. Our results can be viewed as generalizing the
MAX CUT approximation algorithm of Goemans and Williamson to the
problem of finding an approximate CUT COVER. In fact, our techniques
also lead to improved approximations for the MAX $k$-CUT
problem~\cite{FriJer}.  We also establish a duality relationship
between the value of the optimum solution to our semidefinite program
and the Lov\'asz
$\vartheta$-function~\cite{GLS,glands,Lovasz:Shannon}.  We show lower
bounds on the gap between the optimum solution of our semidefinite
program and the actual chromatic number; by duality this also
demonstrates interesting new facts about the $\vartheta$-function.

Alon and Kahale~\cite{alon} use related techniques to devise a
polynomial time algorithm for 3-coloring random graphs drawn
 from a ``hard'' distribution on the space of all 3-colorable graphs.
Recently, Frieze and Jerrum~\cite{FriJer} have used a semidefinite
programming formulation and randomized rounding strategy essentially
the same as ours to obtain improved approximations for the MAX $k$-CUT
problem with large values of $k$. Their results required a more
sophisticated version of our analysis, but for the coloring problem
our results are tight up to poly-logarithmic factors and their
analysis does not help to improve our bounds.

Semidefinite programming relaxations are an extension of the linear
programming relaxation approach to approximately solving NP-complete
problems.  We thus present our work in the style of the classical
LP-relaxation approach.  We begin in Section~\ref{sec:relaxation} by
defining a relaxed version of the coloring problem.  Since we use a
more complex relaxation than standard linear programming, we must show
that the relaxed problem can be solved; this is done in
Section~\ref{sec:solution}.  We then show relationships between the
relaxation and the original problem.  In Section~\ref{sec:bound}, we
show that (in a sense to be defined later) the value of the relaxation
bounds the value of the original problem.  Then, in
Sections~\ref{sec:semicoloring},~\ref{sec:plane},~and~\ref{sec:rounding},
we show how a solution to the relaxation can be
``rounded'' to make it a solution to the original problem.  Combining
the last two arguments shows that we can find a good approximation.
Section~\ref{sec:solution}, Section~\ref{sec:bound}, and
Sections~\ref{sec:semicoloring}--\ref{sec:rounding} are in fact
independent and can be read in any order after the definitions in
Section~\ref{sec:relaxation}.  In Section~\ref{sec:duality}, we
investigate the relationship between our fractional relaxations 
and the Lov\'asz
$\vartheta$-function, showing that they are in fact dual to one
another.  We investigate the approximation error inherent in our
formulation of the chromatic number via semi-definite programming in
Section~\ref{sec:tightness}.

\section{A Vector Relaxation of Coloring}
\label{sec:relaxation}

In this section, we describe the relaxed coloring problem whose
solution is in turn used to approximate the solution to the coloring
problem.  Instead of assigning colors to the vertices of a graph, we
consider assigning ($n$-dimensional) unit {\em vectors} to the vertices.
To capture the property of a coloring, we aim for the vectors of
adjacent vertices to be ``different'' in a natural way.  The {\em
vector $k$-coloring} that we define plays the role that a hypothetical
``fractional $k$-coloring'' would play in a classical
linear-programming relaxation approach to the problem.  Our relaxation
is related to the concept of an orthonormal representation of a
graph~\cite{Lovasz:Shannon,GLS}.

\begin{definition}
Given a graph $G =(V,E)$ on $n$ vertices, and a real number $k \geq 1$,
a {\em vector $k$-coloring}
of $G$ is an assignment of unit vectors $v_i$ from the space $\Re^n$
to each vertex $i \in V$, such that for any two adjacent vertices $i$
and $j$ the dot product of their vectors satisfies the inequality
$$\dotprod{v_i}{v_j} \leq -{1 \over {k-1}}.$$
\end{definition}

The definition of an {\em orthonormal
representation}~\cite{Lovasz:Shannon,GLS} requires that the given dot
products be equal to zero, a weaker requirement than the one above.

\section{Solving the Vector Coloring Problem}
\label{sec:solution}

In this section we show how the vector coloring relaxation can be solved
using semidefinite programming.  The methods in this section closely
mimic those of Goemans and Williamson~\cite{GW:maxcut}.

To solve the problem, we need the following auxiliary definition.

\begin{definition}
Given a graph $G = (V,E)$ on $n$ vertices, a {\em matrix $k$-coloring}
of the graph is an $n \times n$ symmetric positive semidefinite matrix
$M$, with $m_{ii}=1$ and $m_{ij} \leq -{1 / (k-1)}$ if $\{i,j\} \in
E$.
\end{definition}

We now observe that matrix and vector $k$-colorings are in fact
equivalent (cf.~\cite{GW:maxcut}).  Thus, to solve the vector coloring
relaxation it will suffice to find a matrix $k$-coloring.

\begin{fact}
A graph has a vector $k$-coloring if and only if it has matrix
$k$-coloring.  Moreover, a vector $(k + \epsilon)$-coloring can be
constructed from a matrix $k$-coloring in time polynomial in $n$ and
$\log(1/\epsilon)$.
\end{fact}
Note that an exact solution cannot be found, as some of the
values in it may be irrational.
\begin{proof}
Given a vector $k$-coloring $\{v_i\}$, the matrix $k$-coloring is
defined by $m_{ij} = \dotprod{v_i}{v_j}$.  For the other direction, it
is well known that for every symmetric positive definite matrix $M$
there exists a square matrix $U$ such that $UU^T = M$ (where $U^T$ is
the transpose of $U$).
The rows of $U$ are
vectors $\{u_i\}_{i=1}^n$ that form a vector $k$-coloring of $G$.

An $\delta$-close approximation to the matrix
$U$ can be found in time polynomial in $n$ and $\log(1/\delta)$
 using the {\em Incomplete Cholesky
Decomposition}~\cite{GW:maxcut,Golub:Matrix}.
(Here by $\delta$-close we mean a matrix $U'$ such that $U'U'^T -
M$ has $L_\infty$ norm less than $\delta$.)
This in turn gives a vector $(k+\epsilon)$-coloring of the graph,
provided $\delta$ is chosen appropriately.
\end{proof}

\begin{lemma}
\label{lem:findit}
If a graph $G$ has a vector $k$-coloring then a vector $(k +
\epsilon)$-coloring of the graph can be constructed in time polynomial
in $k$, $n$, and $\log (1/ \epsilon)$.
\end{lemma}
\begin{proof}
Our proof is similar to those of Lov\'asz~\cite{Lovasz:Shannon} and
Goemans-Williamson~\cite{GW:maxcut}. We construct a semidefinite
optimization problem (SDP) whose optimum is $-{1 / (k-1)}$ when
$k$ is the smallest real number such that a
matrix
$k$-coloring of $G$
exists. The optimum solution also provides a
matrix
$k$-coloring of
$G$.
\[
\begin{array}{ll}
\mbox{minimize} & {\displaystyle
\alpha } \\[2ex]
\mbox{where} &\mbox{$\{m_{ij}\}$ is positive semidefinite}\\[1ex]
\mbox{subject to}  &
        \begin{array}[t]{rcl}
\displaystyle   m_{ij} &\leq &\displaystyle \alpha ~~~~ \mbox{if $(i,j)
\in E$}\\[1ex]
\displaystyle   m_{ij} &= &m_{ji} \\[1ex]
\displaystyle   m_{ii} &= &1.
        \end{array}
\end{array}
\]
Consider a graph which has a vector
(and matrix)
$k$-coloring.  This means there is a solution to the above
semidefinite program with $\alpha = -1 / (k-1)$.  The ellipsoid
method or other interior point based methods~\cite{GLS,Alizadeh} can
be employed to find a feasible solution where the value of the
objective is at most ${-1 / (k-1)} + \delta$ in time
polynomial in $n$ and $\log{1 / \delta}$. This implies that for
all $\{i,j\} \in E$, $m_{ij}$ is at most $\delta - {1 / (k-1)}$,
which is at most ${-1 / (k + \epsilon - 1)}$ for $\epsilon =
2\delta (k-1)^2$, provided $\delta \leq {1 / 2(k-1)}$.
Thus a
matrix $(k +\epsilon)$-coloring can be found in time polynomial in
$k$, $n$ and $\log ( 1 / \epsilon)$.  From the matrix coloring, the
vector coloring can be found in polynomial time
 as was noted in the
previous lemma
\end{proof}

For the remainder of the paper, we will ignore the $\epsilon$ error
term of Lemma~\ref{lem:findit} since it can be made so small as to be
irrelvant to our analysis.

\section{Relating the Original and Relaxed Solutions}
\label{sec:bound}

In this section, we show that our vector coloring problem is a useful
relaxation because the solution to it is related to the solution of
the original problem.  In order to understand the quality of the
relaxed solution, we need the following geometric lemma:

\begin{lemma}
\label{kcolourable}
For all positive integers $k$ and $n$ such that $k \leq n+1$,
there exist $k$ unit vectors in $\Re^n$ such that the dot product of
any distinct pair is $-1/ (k-1)$.
\end{lemma}
\begin{proof}
  Clearly it suffices to prove the lemma for $n=k-1$. (For other
  values of $n$, we make the coordinates of the vectors $0$ in all but
  the first $k-1$ coordinates.) We begin by proving the claim for
  $n=k$.  We explicitly provide unit vectors $v_1^{(k)},\ldots,
  v_k^{(k)} \in \Re^{k-1}$ such that
  $\dotprod{v_i^{(k)}}{v_j^{(k)}}\leq {-1 / (k-1)}$ for $i \ne j$.
  The vector $v_i^{(k)}$ is $-\sqrt{1 \over k(k-1)}$ in all
  coordinates except the $i$th coordinate. In the $i$th coordinate
  $v_i^{(k)}$ is $\sqrt{k-1 \over k}$. It is easy to verify that the
  vectors are unit length and that their dot products are exactly
  $-1\over k-1$.
  
  As given, the vectors are in a $k$-dimensional space.  Note,
  however, that the dot product of each vector with the all-1's
  vector is 0.  This shows that all k of the vectors are actually in a
  (k-1)-dimensional hyperplane of the k-dimensional space.  This
  proves the lemma.
\end{proof}

\begin{corollary}
Every $k$-colorable graph $G$ has a vector $k$-coloring.
\end{corollary}
\begin{proof}
Bijectively map the $k$ colors to the $k$ vectors defined in the
previous lemma.
\end{proof}

Note that a graph is vector $2$-colorable if and only if it is
$2$-colorable.  Lemma~\ref{kcolourable} is tight in that it provides
the best possible value for minimizing the maximum dot-product among $k$
unit vectors.  This can be seen from the following lemma.

\begin{lemma}
\label{ngbd}
Let $G$ be vector $k$-colorable and let $i$ be a vertex in $G$. The
induced subgraph on the neighbors of $i$ is vector $(k-1)$-colorable.
\end{lemma}
\begin{proof}
Let $v_1,\ldots,v_n$ be a vector $k$-coloring of $G$ and assume
without loss of generality that $v_i = (1,0,0,\ldots,0)$.
Associate with each neighbor $j$ of $i$ a vector $v'_j$ obtained by
projecting $v_j$ onto coordinates $2$ through $n$ and then scaling it
up so that $v'_j$ has unit length.  It suffices to show that for any
two adjacent vertices $j$ and $j'$ in the neighborhood of $i$,
$\dotprod{v'_j} {v'_{j'}} \leq {-1 / (k-2)}$.

Observe first that the projection of $v_j$ onto the first coordinate
is negative and has magnitude at least $1 /(k-1)$.  This implies that the
scaling factor for $v'_j$ is at least $k-1 \over \sqrt{k(k-2)}$.  Thus,
$$\dotprod{v'_j}{v'_{j'}} \leq {(k-1)^2 \over k(k-2)} ( \dotprod{v_{j}}
{v_{j'}} - {1 \over (k-1)^2} ) \leq {-1 \over {k-2}}.$$
\end{proof}

A simple induction using the above lemma shows that any graph
containing a $(k+1)$-clique is not $k$-vector colorable.  Thus the
``vector chromatic number'' lies between between the clique and
chromatic number.  This also shows that the analysis of
Lemma~\ref{kcolourable} is tight in that $-{1 / (k-1)}$ is the
minimum possible value of the maximum of the dot-products of $k$
vectors.

In the next few sections we prove the harder part, namely, if a graph
has a vector $k$-coloring then it has an $\Olog(\Delta^{1-2/k})$ and
an $\Olog(n^{1 - {3/(k+1)}})$-coloring.

\section{Semicolorings}
\label{sec:semicoloring}

Given the solution to the relaxed problem, our next step is to show how
to ``round'' the solution to the relaxed problem in order to get a
solution to the original problem.  Both of the rounding techniques we
present in the following sections produce the coloring by working
through an almost legal {\em semicoloring} of the graph, as defined
below.

\begin{definition}
A {\em $k$-semicoloring} of a graph $G$ is an assignment of $k$ colors
to  at least half its vertices such that no two adjacent vertices are
assigned the same color.
\end{definition}

\comment{
The following lemma follows immediately.

\begin{lemma}
An assignment of $k$ colors to the vertices of a graph $G$ such that 
at most $|V(G)|/2$ edges are incident on two vertices of the same color
is a $k$-semicoloring of $G$.
\end{lemma}

\begin{proof}
For every edge s.t. both its endpoints are assigned the same color, delete
one of its endpoints from the graph. At the end of this procedure, we
are left with a legally colored graph on at least half the vertices.
\end{proof}
}

An algorithm for semicoloring leads naturally to a
coloring algorithm as shown by the following lemma. The algorithm uses 
up at most a logarithmic factor more colors than the semicoloring
algorithm. Furthermore, we do not even lose this logarithmic factor if
the semicoloring algorithm uses a polynomial number of colors (which
is what we will show we use).

\begin{lemma}
\label{semi vs color lemma}
If an algorithm $A$ can $k_i$-semicolor any $i$-vertex subgraph of
graph $G$ in randomized
polynomial time, where $k_i$ increases with $i$, then $A$
can be used to $O(k_n\log n)$-color $G$.  Furthermore, if there exists
$\epsilon > 0$ such that for all $i$, $k_i =\Omega(i^{\epsilon})$,
then $A$ can be used to color $G$ with $O(k_n)$ colors.
\end{lemma}
\begin{proof}
We show how to construct a coloring algorithm $A'$
to color any subgraph $H$ of $G$. $A'$ starts by using $A$ to semicolor
$H$. Let $S$ be the subset of vertices that have not been assigned
a color by $A$.
Observe that $|S| \leq |V(H)|/2$.  $A'$ fixes
the colors of vertices not in $S$,
and then recursively colors the induced subgraph on $S$ using a new set of
colors.

Let $c_i$ be the maximum number of colors used by $A'$ to color any
$i$-vertex subgraph. Then $c_i$ satisfies the recurrence
$$c_i \leq c_{i/2} + k_i$$
It is easy to see that this any $c_i$ satisfying this recurrence, must
satisfy $c_i \leq k_i \log i$. In particular this implies that $c_n \leq
O(k_n \log n)$. Furthermore for the case where $k_i = \Omega(i^\epsilon)$
the above recurrence is satisfied only when $c_i = \Theta(k_i)$.
\end{proof}

\remove{
Suppose we find a semicoloring.  Since there are at most $n/4$ ``bad''
edges which have among them at most $n/2$ endpoints, there must be at
least $n/2$ vertices with no incident bad edges.  The colors assigned
to these $n/2$ good vertices form a legal coloring of these vertices.
Suppose we fix the colors of these vertices and set them aside.  We
can then find a semicoloring (with new colors) of the graph induced by
the remaining vertices, and again set aside a set of legally colored
vertices.  Since at each iteration we set aside a constant fraction of
the vertices, we will finish after $O(\log n)$ semicoloring steps.
The total number of colors used is $O(\sum_{i=1}^{\lceil \log n \rceil}
k_{2^i})$, which is certainly $O(k_n\log n)$.  The tighter bound for larger
$k$ is obtained from the fact that the total number of colors used above
is a telescoping sum adding up to $O(k_n)$ colors in all.
\end{proof}
}

Using the above lemma, we devote the next two sections to
algorithms for transforming vector colorings into semicolorings.

\section{Rounding via Hyperplane Partitions}
\label{sec:plane}

We now focus our attention on vector $3$-colorable graphs, leaving the
extension to general $k$ for later.  Let $\Delta$ be the maximum
degree in a graph $G$.  In this section, we outline a randomized
rounding scheme for transforming a vector $3$-coloring of $G$ into an
$O(\Delta^{\log_3 2})$-semicoloring, and thus into an
$O(\Delta^{\log_3 2} \log n)$-coloring of $G$.
Combining this method with a technique of Wigderson~\cite{wigderson} 
yields an
$O(n^{0.386})$-coloring of $G$.
The method is based
on that of~\cite{GW:maxcut} and is weaker than the method we describe in the
following section; however, it introduces several of the ideas we will
use in the more powerful algorithm.

Assume we are given a vector $3$-coloring $\{v_i\}_{i=1}^n$.  Recall that the
unit vectors $v_i$ and $v_j$ associated with an adjacent pair of
vertices $i$ and $j$ have a dot product of at most $-1/2$, implying
that the angle between the two vectors is at least $2\pi/3$ radians 
(120 degrees).

\begin{definition}
Consider a hyperplane $H$. We say that $H$ {\em separates} two vectors
if they do not lie on the same side of the hyperplane. For any edge
$\{i,j\} \in E$, we say that the hyperplane $H$ {\em cuts} the edge if
it separates the vectors $v_i$ and $v_j$.
\end{definition}

In the sequel, we use the term {\em random hyperplane} to denote the
unique hyperplane containing the origin and having as its normal a
random unit vector $v$ uniformly distributed on the unit sphere $S_n$.
The following lemma is a restatement of Lemma 1.2 of
Goemans-Williamson~\cite{GW:maxcut}.

\begin{lemma}[Goemans-Williamson~\cite{GW:maxcut}]
Given two vectors at an angle of $\theta$, the probability that they
are separated by a random hyperplane is exactly $\theta/\pi$.
\end{lemma}

We conclude that given a vector 3-coloring, for any edge $\{i,j\} \in
E$, the probability that a random hyperplane cuts the edge is exactly
$2/3$. It follows that the expected fraction of the edges in $G$ that
are cut by a random hyperplane is exactly $2/3$.  Suppose that we pick
$r$ random hyperplanes independently. Then, the probability that an
edge is not cut by one of these hyperplanes is $(1/3)^r$, and the
expected fraction of the edges not cut is also $(1/3)^r$.

We claim that this gives us a good semicoloring algorithm for the
graph $G$.  Notice that $r$ hyperplanes can partition $\Re^n$ into at
most $2^r$ distinct regions. (For $r \leq n$ this is tight since $r$
hyperplanes can create exactly $2^r$ regions.) An edge is cut by one of
these $r$ hyperplanes if and only if the vectors associated with its
end-points lie in distinct regions. Thus, we can associate a distinct
color with each of the $2^r$ regions and give each vertex the color of
the region containing its vector.  The expected number of edges whose
end-points have the same color is $(1/3)^r m$, where $m$ is the number
of edges in $E$.

\begin{theorem}
\label{thm:naive}
If a graph has a vector $3$-coloring, then it has an $O (\Delta^{\log_3
2})$-semicoloring that can be constructed from the vector $3$-coloring
in polynomial time with high probability.
\end{theorem}
\begin{proof}
We use the random hyperplane method just described.  Fix $r = 2+\lceil \log_3
\Delta \rceil$, and note that $(1/3)^r \leq 1/9\Delta$ and that $2^r =
O (\Delta^{\log_3 2})$.  As noted above, $r$ hyperplanes chosen
independently at random will cut an edge with probability $1-1/9\Delta$.
Thus the expected number of edges that are not cut is $m/9\Delta \le
n/18 < n/8$, since the number of edges is at most $n\Delta/2$.  By
Markov's inequality (cf. \cite{MoRa}, page 46), 
the probability that the number of uncut edges is
more than twice the expected value is at most $1/2$.  Thus, with
probability at least 1/2 we get a coloring with at most $n/4$ uncut
edges.  Delete one endpoint of each such edge leaves a set of $3n/4$
colored vertices with no uncut edges---ie, a semicoloring.

Repeating the entire process $t$ times means that we will find a
$O(\Delta^{\log_3 2})$-semicoloring with probability at
least $1 - 1/2^t$.
\end{proof}

Noting that $\log_3 2 < 0.631$ and that $\Delta \leq n$, this theorem
and Lemma~\ref{semi vs color lemma} implies a semicoloring using
$O(n^{0.631})$ colors. 

By varying the number of hyperplanes, we can arrange for a tradeoff
between the number of colors used and the number of edges that
violate the resulting coloring.  This may be useful in some
applications where a nearly legal coloring is good enough.

\subsection{Wigderson's Algorithm}

Our coloring can be improved using the following idea due to
Wigderson~\cite{wigderson}.  Fix a threshold value $\delta$. If there
exists a vertex of degree greater than $\delta$, pick any one such
vertex and 2-color its neighbors (its neighborhood is vector
2-colorable and hence 2-colorable).  The colored vertices are removed
and their colors are not used again.  Repeating this as often as
possible (or until half the vertices are colored) brings the maximum
degree below $\delta$ at the cost of using at most $2n/\delta$ colors.
At this point, we can semilcolor the remainder with
$O(\delta^{0.631})$ colors.  Thus, we can obtain a semicoloring using
$O(n/\delta + \delta^{0.631})$ colors.  The optimum choice of $\delta$
is around $n^{0.613}$, which implies a semicoloring using
$O(n^{0.387})$ colors.  This semicoloring can be used to legally color
$G$ using $O(n^{0.387})$ colors by applying Lemma~\ref{semi vs color
  lemma}.

\begin{corollary}
\label{cor:naive}
A 3-colorable graph with $n$ vertices can be colored using $O(n^{0.387}
)$ colors by a polynomial time randomized algorithm.
\end{corollary}

The bound just described is (marginally) weaker than the guarantee of
a $O(n^{0.375})$ coloring due to Blum~\cite{blum94}.  We now improve
this result by constructing a semicoloring with fewer colors.

\section{Rounding via Vector Projections}
\label{sec:rounding}

In this section we start by proving the following more powerful
version of Theorem~\ref{thm:naive}. A simple application of
Wigderson's technique to this algorithm yields our final coloring
algorithm.

\begin{theorem}
\label{thm:pre-wig}
For every integer function $k = k(n)$, a vector $k$-colorable graph 
with maximum degree $\Delta$
can be semi-colored
with at most $O(\Delta^{1-2/k} \sqrt{\ln \Delta})$ colors in 
probabilistic polynomial time.
\end{theorem}

As in the previous section, this has immediate consequences for
approximate coloring. 

To prove Theorem~\ref{thm:pre-wig}, given a vector $k$-coloring, we
show that it is possible to extract an independent set of size
$\Omega(n/(\Delta^{1-2/k}\sqrt{\ln \Delta}))$.  If we assign one color
to this set and recurse on the rest of the vertices, we will end up
using $O(\Delta^{1-2/k}\sqrt{\ln \Delta})$ colors in all to assign
colors to half the vertices and the result follows. To find such a
large independent set, we give a randomized procedure for selecting an
induced subgraph with $n'$ vertices and $m'$ edges such that
$E[n'-m']= \Omega(n/(\Delta^{1-2/k} \sqrt{\ln \Delta}))$.  It follows
that with a polynomial number of repeated trials, we have a high
probability of choosing a subgraph with $n'-m'=
\Omega(n/(\Delta^{1-2/k} \sqrt{\ln \Delta}))$.  Given such a graph, we
can delete one endpoint of each edge, leaving an independent set of
size $n'-m'= \Omega(n/(\Delta^{1-2/k} \sqrt{\ln \Delta}))$, as
desired.

We now give the details of the construction.  Suppose we have a vector
$k$-coloring assigning unit vectors $v_i$ to the vertices.  We fix
a parameter $c=c_{k,\Delta}$ to be specified later.  
We choose a random $n$-dimensional
vector $r$ according to a distribution to be specified soon.
The subgraph consists of all 
vertices $i$ with $v_i \cdot r \ge c$.  Intuitively, since endpoints
of an edge have vectors pointing away from each other, if the vector
associated with a vertex has a large dot product with $r$, then the
vector corresponding to an adjacent vertex will not have such a large
dot product with $r$ and hence will not be selected. Thus, only a few
edges are likely to be in the induced subgraph on the selected set of
vertices.

To complete the specification of this algorithm and to analyze it, 
we need some basic facts about some probability distributions in $\Re^n$.

\subsection{Probability Distributions in $\Re^n$}

Recall that the {\em standard normal distribution} has the density
function $\phi(x) = \frac{1}{\sqrt{2\pi}} e^{-x^2/2}$ with
distribution function $\Phi(x)$, mean 0, and variance 1. A random
vector $r = (r_1,\ldots,r_n)$ is said to have the {\em $n$-dimensional
standard normal distribution} if the components $r_i$ are independent
random variables, each component having the standard normal
distribution. It is easy to verify that this distribution is
spherically symmetric, in that the direction specified by the vector
$r$ is uniformly distributed. (Refer to Feller~\cite[v.~II]{feller},
Knuth~\cite[v.~2]{knuth}, and R\'enyi~\cite{renyi} for further details
about the higher dimensional normal distribution.)

Subsequently, the phrase ``random $d$-dimensional vector'' will always
denote a vector chosen from the $d$-dimensional standard normal
distribution.  A crucial property of the normal distribution which
motivates its use in our algorithm is the following theorem
paraphrased from R\'enyi~\cite{renyi} (see also Section III.4 of
Feller~\cite[v.~II]{feller}).

\begin{theorem}[Theorem IV.16.3~\cite{renyi}]
Let $r = (r_1, \ldots, r_n)$ be a random $n$-dimensional vector.  The
projections of $r$ onto two lines ${\ell}_1$ and ${\ell}_2$ are
independent (and normally distributed) if and only if ${\ell}_1$ and
${\ell}_2$ are orthogonal.
\end{theorem}

Alternatively, we can say that under any rotation of the coordinate
axes, the projections of $r$ along these axes are independent standard
normal variables. In fact, it is known that the only distribution with
this strong spherical symmetry property is the $n$-dimensional
standard normal distribution. The latter fact is precisely the reason
behind this choice of distribution\footnote{Readers familiar with
physics will see the connection to Maxwell's law on the distribution of
velocities of molecules in $\Re^3$. Maxwell started with the assumption
that in {\em every} Cartesian coordinate system in $\Re^3$, the three
components of the velocity vector are mutually independent and had
expectation zero. Applying this assumption to rotations of the axes,
we conclude that the velocity components must be independent normal
variables with identical variance. This immediately implies Maxwell's
distribution on the velocities.}
in our algorithm. In particular, we will make use of the following
corollary to the preceding theorem.

\begin{corollary}
\label{cor:first}
Let $u$ be any unit vector in $\Re^n$.
Let $r = (r_1, \ldots, r_n)$ be a random vector (of i.i.d.~standard
normal variables).  
The projection of $r$ along $u$, 
given by dot product $\dotprod{u}{r}$, is distributed according to 
the standard ($1$-dimensional) normal distribution.
\end{corollary}

It turns out that even if $r$ is a random $n$-dimensional {\em unit}
vector, the above corollary still holds in the limit: as $n$ grows,
the projections of $r$ on orthogonal lines approach (scaled)
independent normal distributions.  Thus using a random unit vectors
for our projection turns out to be equivalent to using random normal
vectors in the limit, but is messier to analyze.

\newcommand{\gauss}{\phi}
\newcommand{\N}{N}

Let $\N(x)$ denote the
tail of the standard normal distribution. I.e.,
\[
\N(x) = \int_x^{\infty} \gauss(y)\, dy.
\]
We will need the following well-known bounds on the tail of
the standard normal distribution.
(See, for instance, Lemma VII.2 of Feller~\cite[v.~I]{feller}.)

\begin{lemma} For every $x>0$,
\label{lem:second}
\[
\gauss(x)\left(\frac1x-\frac1{x^3}\right) < \N(x) < \gauss(x)\cdot\frac1x
\]
\end{lemma}
\begin{proof}
The proof is immediate from inspection of the following equations
relating the three quantities in the desired inequality to integrals
involving $\gauss(x)$, and the fact $\gauss(x)/x$ is finite for every
$x > 0$.
\begin{eqnarray*}
\gauss(x)\left(\frac1x-\frac1{x^3}\right) &= &\int_x^\infty  \gauss(y) \left(1-\frac3{y^4}\right)\, dy,\\
\N(x) &= &\int_x^\infty  \gauss(y) \,dy,\\
\gauss(x) \cdot \frac1x &= &\int_x^\infty  \gauss(y)\left(1+\frac1{y^2}\right)\, dy.\\
\end{eqnarray*}
\end{proof}

\subsection{The Analysis}

We are now ready to complete the specification of the coloring
algorithm.  Recall that our goal is to repeatedly strip away
large independent sets from the graph.  We actually set an
easier intermediate goal: find an induced subgraph with a large number
$n'$ of vertices and a number $m' \ll n'$ of edges.  Given such a
graph, we can delete one endpoint of each edge to leave an independent
set on $n'-m'$ vertices that can be colored and removed.

As discussed above, to find this sparse graph, we choose a random
vector $r$ and take all vertices whose dot product with $r$ exceeds a
certain value $c$.  Let the induced subgraph on these vertices have
$n'$ vertices and $m'$ edges.  We show that for sufficiently large
$c$, $n' \gg m'$ and we get an independent set of size roughly $n'$.
Intuitively, this is true for the following reason.  Any particular
vertex has some particular probability $p=p(c)$ of being near $r$
and thus being ``captured'' into our set.  However, if two vertices
are adjacent, the probability that they {\em both} land near $r$ is
quite small because the vector coloring has placed them far apart.

For example, in the case of 3-coloring, when the probability that a
vertex is captured is $p$, the probability that both endpoints of an
edge are captured is roughly $p^4$ (this is counter the intuition that
the probability should go as $p^2$, and follows from the fact that we
force adjacent vertices to be far apart---see below).  It follows that
we end up capturing (in expectation) a set of $pn$ vertices that
contains (in expectation) only $p^4 m < p^4 \Delta n$ edges in a
degree-$\Delta$ graph.  In such a set, at least $pn-p^4 \Delta n$ of
the vertices have no incident edges, and thus form an independent set.
We would like this independent set to be large.  Clearly, we need to
make $p$ small enough to ensure $p^4 \Delta n \ll pn$, meaning $p \ll
\Delta^{-1/3}$.  Taking $p$ much smaller only decreases the size of
the independent set, so it turns out that our best choice is to take
$p \approx \Delta^{-1/3}/2$, yielding an indpendent set of size
$\Omega(n\Delta^{-1/3})$.  Repeating this capture process many times
therefore achieves an $\tilde{O}(\Delta^{1/3})$ coloring.

We now formalize the intuitive argument.  The vector $r$ will be a
random $n$-dimensional vector.  We precisely compute the expectation
of $n'$, the number of vertices captured, and the expectation of $m'$,
the number of edges in the induced graph of the captured vertices.  We
first show that when $r$ is a random normal vector and our projection
threshold is $c$, the expectation of $n'-m'$ exceeds $n(N(c)-\Delta
N(ac))$ for a certain constant $a$ depending on the vector chromatic
number. We also show that $\N(ac)$ grows roughly as $\N(c)^{a^2}$.
(For the case of 3-coloring we have $a=2$, 
and thus if $\N(c) = p$, then $\N(ac) \approx p^4$.)
By picking a sufficiently large $c$,
we can find an independent set of size
$\Omega(\N(c))$. 
(In the following lemma, $n'$ and $m'$ are 
functions of $c$: we do not make this dependence explicit.)

\begin{lemma}
Let $a=\sqrt{\frac{2(k-1)}{k-2}}$.  Then for any $c$,
\[
E[n'-m'] > n\left(N(c)-\frac{\Delta N(ac)}{2}\right).
\]
\end{lemma}
\begin{proof}
  We first bound $\E{n'}$ from below.  Consider a particular vertex
  $i$ with assigned vector $v_i$.  The probability that it is in the
  selected set is just $\Pr{v_i \cdot r \ge c}$.  By
  Corollary~\ref{cor:first}, $v_i \cdot r$ is normally distributed and
  thus this probability is $\N(c)$. By linearity of expectations, the
  expected number of selected vertices $\E{n'} = n\N(c)$.

  Now we bound $\E{m'}$ from above.  Consider an edge with endpoint
  vectors $v_1$ and $v_2$.  The probability that this edge is in the
  induced subgraph is the probability that both endpoints are
  selected, which is
\begin{eqnarray*}
\Pr{v_1 \cdot r \ge c \mbox{ and }v_2 \cdot r \ge c }
&\le &
\Pr{(v_1 +v_2) \cdot r \ge 2c }\\
&= &
\Pr{\frac{v_1+v_2}{\|v_1+v_2\|} \cdot r \ge \frac{2c}{\|v_1+v_2\|}}\\
&= &
\N\left(\frac{2c}{\|v_1+v_2\|}\right),
\end{eqnarray*}
where the expression follows from Corollary~\ref{cor:first} applied to the
preceding probability expression. We now observe that
\begin{eqnarray*}
\|v_1+v_2\| &=
&\sqrt{v_1^2+v_2^2+2v_1\cdot v_2}\\
&\leq &\sqrt{2-2/(k-1)}\\
&= &\sqrt{2(k-2)/(k-1)}\\
&= &2/a.
\end{eqnarray*}
It follows that the probability that both endpoints of an edge are
selected is at most $\N(ac)$.  
If the graph has maximum degree
$\Delta$, then the total number of edges is at most $n\Delta/2$.
Thus the expected number of selected edges, $\E{m'}$, is at most
$n\Delta \N(ac)/2$.

Combining the previous arguments, we deduce that 
\[
\E{n'-m'} \geq n\N(c)-n\Delta \N(ac)/2.
\]
\end{proof}

We now determine a $c$ such that $\Delta \N(ac) < \N(c)$.  This
will give us an expectation of at least $N(c)/2$ in the above lemma.
Using the bounds on $\N(x)$ in Lemma~\ref{lem:second}, we find that
\begin{eqnarray*}
\frac{\N(c)}{\N(ac)} &\ge
&\frac{(\frac1c-\frac{1}{c^3})e^{-c^2/2}}{e^{-a^2c^2/2}/ac} \\
&= &a\left(1-\frac1{c^2}\right)e^{(a^2-1)c^2/2} \\
&\ge &\sqrt{2}\left(1-\frac1{c^2}\right)e^{(a^2-1)c^2/2} 
\end{eqnarray*}
The last equation holds since $a = \sqrt{2(k-1)/(k-2)} > \sqrt{2}$.
Thus if we choose $c$ so that $1 - 1/c^2 \geq \frac1{\sqrt{2}}$ and 
$e^{(a^2-1)c^2/2} \geq \Delta$, then we get 
$\Delta \N(ac) < \N(c)$. Both conditions are satisfied, for sufficiently
large $\Delta$, if
we set 
\[
c = \sqrt{2\frac{(k-2)}{k} \ln \Delta }.
\]
(For smaller values of $\Delta$ we can use the greedy 
$\Delta+1$-coloring algorithm to get a color the graph with 
a bounded number of colors, where the bound is independent of $n$.)

For this choice of $c$, we find that 
the independent set that is found has size at least
\begin{eqnarray*}
\E{n'-m'} 
        & \geq & n\N(c)/2 \\
        &\geq & \Omega\left(ne^{-c^2/2}\left(\frac1c - \frac1{c^3}\right)\right)\\
        &\geq & \Omega\left(\frac{n}{ \Delta^{1 - \frac2k} {\sqrt{\ln\Delta}} }\right)
\end{eqnarray*}
as desired. This concludes the proof of Theorem~\ref{thm:pre-wig}.

\subsection{Adding Wigderson's Technique}

To conclude, we now determine absolute approximation ratios
independent of $\Delta$. This involves another application
of Wigderson's technique. If the graph has any vertex of large
degree, then we use the fact that its neighborhood is large
and is vector $(k-1)$-chromatic, to find a large independent
set in its neighborhood. If no such vertex exists, then the 
graph has small maximum degree, so we can use Theorem~\ref{thm:pre-wig}
to find a large independent set in the graph. After extracting 
such an independent set, we recurse on the rest of the graph.
The following lemma describes the details, and the correct
choice of the threshold degree.

\comment{
We identify a positive real function
$r(k)$ such that we can color a vector $k$-chromatic graph with
at most $n^{r(k)}$ colors.  For each $k$, we establish a degree
threshold $\Delta_k =
\Delta_k(n)$.  While the degree exceeds $\Delta_k$, we take a
neighborhood of a vertex of degree $d \geq \Delta_k$ and recursively
$d^{r(k - 1)}$-color it and discard it (by Lemma~\ref{ngbd} the
neighborhood is vector $(k - 1)$-chromatic).
The average number of colors used per vertex in this process is
$d^{r(k -1) -1} \leq
\Delta_k^{r(k-1)-1}$.
Thus the total number of colors used up in this process is at most
$n\Delta_k^{r(k-1)-1}$ colors.  Once the degree is less than
$\Delta_k$, we use our coloring algorithm directly to use an
additional $\tilde{O}(\Delta_k^{1-2/k})$ colors. In order to estimate
$\Delta_k$ and $r(k)$ we ignore polylog factors. We balance the colors
used in each part by setting
\[
n\Delta_k^{r(k-1)-1} = \Delta_k^{1-2/k}\\
\]
which implies that
\begin{eqnarray*}
n &= &\Delta_{k}^{2-2/k-r(k-1)},\\
\Delta_k &= &n^{1/(2-2/k-r(k-1))}.
\end{eqnarray*}
We obtain a coloring with $n^{(1-2/k)/(2-2/k-r(k-1))}$
colors,
in other words \[
r(k) = (1-2/k)/(2-2/k-r(k-1)).\]
By substitution, $r(k) = 1-3/(k+1)$.
This gives the parameters for the following lemma. 
}

\begin{lemma}
\label{lem:post-wig}
For every integer function $k=k(n)$, any vector $k$-colorable 
graph on $n$ vertices can be semicolored with 
$O(n^{1 - 3/(k+1)} \log^{1/2} n)$ colors by a probabilistic
polynomial time algorithm.
\end{lemma}

\begin{proof}
Given a vector $k$-colorable graph $G$, we show how to find an 
independent set of size $\Omega(n^{3/(k+1)}/\log^{1/2} n)$ in the
graph. Assume, by induction on $k$, that there exists a constant
$c > 0$ s.t. we can find an independent set of size 
$c i^{3/(k'+1)}/(\log^{1/2} i)$
in any $k'$-vector chromatic graph on $i$ nodes, for $k'< k$. 
We now prove the inductive assertion for $k$. 

Let $\Delta_k = \Delta_k(n) = n^{k/(k+1)}$. 
If $G$ has a vertex of degree greater than $\Delta_k(n)$, then
we find a large independent set in the neighborhood of $G$. By 
Lemma~\ref{ngbd}, the neighborhood is vector $(k-1)$-colorable.
Hence we can find in this neighborhood, an independent set of
size at least $c (\Delta_k)^{3/k} / (\log^{1/2} \Delta_k) \geq
c n^{3/(k+1)} / (\log^{1/2} n)$.
If $G$ does not have a vertex of degree greater than 
$\Delta_k(n)$, then by Theorem~\ref{thm:pre-wig}, we can find an
independent set of size at least $c n/(\Delta_k)^{1 - 2/k} / \log^{1/2}
\Delta_k \geq c n^{3/(k+1)} / \log^{1/2} n$ in $G$.
This completes the induction.

By now assigning a new color to each such independent set, we find that
we can color at least $n/2$ vertices, using up at most $O(n^{1 - 3/(k+1)}
\log^{1/2} n)$ colors. 
\end{proof}

The semicolorings guaranteed by 
Theorem~\ref{thm:pre-wig}~and~\ref{lem:post-wig} 
can be converted into colorings using Lemma~\ref{semi vs color lemma},
yielding the following theorem.

\begin{theorem}
Any  vector $k$-colorable graph on $n$ nodes with maximum degree $\Delta$
can be colored, in probabilistic polynomial time,  using
$\min\{O(\Delta^{1-2/k}\sqrt{\ln \Delta}\log n), 
O(n^{1-3/(k+1)} \sqrt{\ln n})\}$ colors.
\end{theorem}

\section{Duality Theory}
\label{sec:duality}

The most intensively studied relaxation of a semidefinite programming
formulation to date is the Lov\'asz
$\vartheta$-function~\cite{GLS,glands,Lovasz:Shannon}.  This
relaxation of the clique number of a graph led to the first
polynomial-time algorithm for finding the clique and chromatic numbers
of perfect graphs.  We now investigate a connection between
$\vartheta$ and a close variant of the vector chromatic number.

Intuitively, the clique and coloring problems have a certain
``duality'' since large cliques prevent a graph from being colored
with few colors.  Indeed, it is the equality of the clique and
chromatic numbers in perfect graphs that lets us compute both in
polynomial time.  We proceed to formalize this intuition.  The duality
theory of linear programming has an extension to semidefinite
programming.  With the help of Eva Tardos and David Williamson, we
have shown that in fact the $\vartheta$-function and a close variant
of the vector
chromatic number are semidefinite programming duals to one another and
are therefore equal.

We first define the variant.

\begin{definition}
Given a graph $G=(V,E)$ on $n$ vertices, a {\em strict vector
$k$-coloring} of $G$ is an assignment of unit vectors $u_i$ from the
space $\Re^n$
to each vertex $i \in V$, such that for any two adjacent vertices $i$
and $j$ the dot product of their vectors satisfies the equality
$$\dotprod{u_i}{u_j} = -{1 \over {k-1}}.$$
\end{definition}

As usual we say that a graph is strictly vector $k$-colorable if it has a
strict vector $k$-coloring. The strict vector chromatic number of a
graph is the smallest real number $k$ for which it has a strict
vector $k$-coloring. It follows from the definition that the strict
vector chromatic number of any graph is lower bounded by the vector
chromatic number.

\begin{theorem}
The strict vector chromatic number of $G$ is equal to
$\vartheta(\overline{G})$.
\end{theorem}
\begin{proof}
The dual of our strict vector coloring semidefinite program is as
follows (cf.~\cite{Alizadeh}):
\[
\begin{array}{ll}
\mbox{maximize} & {\displaystyle
- \sum p_{ii} } \\[2ex]
\mbox{where} &\mbox{$\{p_{ij}\}$ is positive semidefinite}\\[1ex]
\mbox{subject to}  &
        \begin{array}[t]{rcl}
\displaystyle   \sum_{i \ne j} p_{ij} &\geq &1 \\[1ex]
\displaystyle   p_{ij} &= &p_{ji} \\[1ex]
\displaystyle   p_{ij} &= &0 \quad \mbox{for $(i,j) \notin E$ and $i\ne j$}
        \end{array}
\end{array}
\]
By duality, the value of this SDP is $-1/(k-1)$ where $k$ is the
strict vector chromatic number.  Our goal is to prove $k=\vartheta$.  As
before, the fact that $\{p_{ij}\}$ is positive semidefinite means we
can find vectors $v_i$ such that $p_{ij} = \dotprod{v_i}{v_j}$. The
last constraint says that the vectors $v$ form an {\em orthogonal
labeling}~\cite{glands}, i.e. that $\dotprod{v_i}{v_j}=0$ for
$(i,j)\notin E$.
We now claim that the above optimization problem can be reformulated as follows:
\[
\mbox{maximize }
\frac{-\sum \dotprod{v_i}{v_i}}{\sum_{i \ne j} \dotprod{v_i}{v_j}}
\]
over all orthogonal labelings $\{v_i\}$.  To see this, consider an
orthogonal labeling and define $\mu = \sum_{i \ne j}
\dotprod{v_i}{v_j}$.  Note this is the value of the first constraint
in the first formulation of the dual (that is, the constraint is $\mu
\le 1$) and of the denominator in the second formulation.  Then in an
optimum solution to the first formulation, we must have $\mu=1$, since
otherwise we can divide each $v_i$ by $\sqrt{\mu}$ and get a feasible
solution with a larger objective value.  Thus the optimum of the
second formulation is at least as large as that of the first.
Similarly, given any optimum $\{v_i\}$ for the second formulation,
$v_i/\sqrt{\mu}$ forms a feasible solution to the first formulation
with the same value.  Thus the optima are equal.  We now manipulate
the second formulation.
\begin{eqnarray*}
\max \frac{-\sum \dotprod{v_i}{v_i}}{\sum_{i \ne j} \dotprod{v_i}{v_j}}
&= &
\max \frac{-\sum \dotprod{v_i}{v_i}}{\sum_{i,j} \dotprod{v_i}{v_j}-\sum
\dotprod{v_i}{v_i}}\\
&= &
\left(\min \frac{\sum_{i,j} \dotprod{v_i}{v_j}-\sum
\dotprod{v_i}{v_i}}{-\sum \dotprod{v_i}{v_i}}\right)^{-1}\\
&= &
\left(\min -\frac{\sum_{i,j} \dotprod{v_i}{v_j}}{\sum
\dotprod{v_i}{v_i}} + 1\right)^{-1}\\
&= &
-\left(\max \frac{\sum_{i,j} \dotprod{v_i}{v_j}}{\sum
\dotprod{v_i}{v_i}} - 1\right)^{-1}.
\end{eqnarray*}
It follows from the last equation that the strict vector chromatic number is
\[
\max \frac{\sum_{i,j} \dotprod{v_i}{v_j}}{\sum \dotprod{v_i}{v_i}}.
\]
However, by the same argument as was used to reformulate the dual, this is
equal to problem of maximizing $\sum_{i,j} \dotprod{v_i}{v_j}$ over
all orthogonal labelings such that $\sum \dotprod{v_i}{v_i} \le 1$.
This is simply Lov\'asz's $\vartheta_3$ formulation of the
$\vartheta$-function~\cite[page 287]{glands}.
\end{proof}

\section{The Gap between Vector Colorings and Chromatic Numbers}
\label{sec:tightness}

The performance of our randomized rounding approach seems far from
optimum.  In this section we ask why, and show that the problem is not
in the randomized rounding but in the gap between the original problem
and its relaxation.  We investigate the following question: given a
vector $k$-colorable graph $G$, how large can its chromatic number be
in terms of $k$ and $n$? We will show that a graph with chromatic
number $n^{\Omega(1)}$ can have bounded vector chromatic number. This
implies that our technique is tight in that it is not possible to
guarantee a coloring with $n^{o(1)}$ colors on all vector 3-colorable
graphs.
\remove{
Lov\'asz~\cite{Lovasz:Personal} pointed out that for a random
graph $\chi = n/\log n$ while $\vartheta = \sqrt{n}$, and that a graph
constructed by Koniagin has $\chi \geq n/2$ and $\vartheta = n^{1/3}$.
However, such large gaps are not known for the case of bounded
$\vartheta$. Our ``bad'' graphs are the so-called Kneser
graphs~\cite{kneser}. (Independent of our results, Szegedy~\cite{Szegedy} has also shown that a similar construction yields
graphs with vector chromatic number at most $3$ that are not
colorable using $n^{0.05}$ colors. Notice that the exponent obtained
 from his result is better than the one shown below.)
}

\begin{definition}
The Kneser graph $K(m,r,t)$ is defined as follows: the vertices are all
possible $r$-sets from a universe of size $m$; and, the vertices $v_i$
and $v_j$ are adjacent if and only if the corresponding $r$-sets satisfy
$|S_i \cap S_j| < t$.
\end{definition}

We will need following theorem of Milner~\cite{milner} regarding
intersecting hypergraphs. Recall that a collection of sets is called
an antichain if no set in the collection contains another.
\begin{theorem}[Milner]
Let $S_1$, $\ldots$, $S_{\alpha}$ be an antichain of sets
 from a universe of size $m$ such that, for all $i$ and $j$,
$$|S_i \cap S_j| \geq t.$$
Then, it must be the case that
$$\alpha \leq {m \choose {\frac{m+t+1}{2}}}.$$
\end{theorem}
Notice that using all $q$-sets, for
$q=(m+t+1)/2$, gives a tight example for this theorem.

The following theorem establishes that the Kneser graphs have a large
gap between their vector chromatic number and chromatic numbers.
\begin{theorem}
Let $n = {m \choose r}$ denote the number of vertices of the graph
$K(m,r,t)$. For $r = m/2$ and $t = m/8$, the graph $K(m,r,t)$ is vector
3-colorable but has chromatic number at least $n^{0.0113}$.
\end{theorem}
\begin{proof}
We prove a lower bound on the Kneser graph's chromatic number $\chi$
by establishing an upper bound on its independence number $\alpha$.
It is easy to verify that the $\alpha$ in Milner's theorem is exactly
the independence number of the Kneser graph.  To bound $\chi$
observe that 
\begin{eqnarray*}
\chi  & \geq & \frac{n}{\alpha} \\
      & \geq & \frac{{m \choose r}}{{m \choose {(m+t)/2}}} \\
      &   =  & \frac{{m \choose m/2}}{{m \choose 9m/16}} \\
      & \approx & \frac{ 2^m(1 - o(1))}{2^{ (1 - o(1))m ((9/16) \lg (16/9) + (7/16) \lg
      (16/7))}} \\
      & \geq & 2^{.0113 m} \mbox{~~~for large enough $m$.}
\end{eqnarray*}
In the above sequence, the fourth line uses the approximation 
\[
{m
\choose \beta m} \approx 2^{m (-\beta \lg \beta - (1 - \beta) \lg
(1-\beta))}/\sqrt{c_{\beta}m},
\]
for every $\beta \in (0,1)$, where
$c_{\beta}$ is a constant depending only on $\beta$.
Using the inequality
\[
n = {m \choose r} \leq 2^m,
\]
we obtain $m \geq \lg n$ and thus
\[
\chi \geq (1.007864) ^ {\lg n} = n^{ \lg 1.007864} \approx n^{0.0113}.
\]

Finally, it remains to show that the vector chromatic number of this
graph is 3. This follows by associating with each vertex $v_i$ an
$m$-dimensional vector obtained from the characteristic vector of the
set $S_i$. In the characteristic vector, $+1$ represents an element
present in $S_i$ and $-1$ represents elements absent from $S_i$. The
vector associated with a vertex is the characteristic vector of $S_i$
scaled down by a factor of $\sqrt{m}$ to obtain a unit vector. Given 
vectors corresponding to sets $S_i$ and $S_j$, the dot product gets a 
contribution of
$-1/m$ for coordinates in $S_i \Delta S_j$ and $+1/m$ for the others.
(Here $A \Delta B$ represents the symmetric difference of the two sets,
i.e., the set of elements that occur in exactly one of $A$ or $B$.)
Thus the dot product of two adjacent vertices,
or sets with
intersection at most $t$, is given by 
$$1 - {2|S_i \Delta S_j| \over m} = 1 - {2 (|S_i| + |S_j| - 2 |S_i \cap
S_j|) \over m} \leq 1 - \frac{4r-4t}{m} = -1/2.$$ 
This implies that the vector chromatic
number is 3.
\end{proof}

More refined calculations can be used to improve this bound somewhat.
\iflong\else
The basic idea is to improve the bound on the vector chromatic number
of the Kneser graph using an appropriately weighted version of the
characteristic vectors.
\fi

\begin{theorem}
There exists a Kneser graph $K(m,r,t)$ that is 3-vector colorable but
has chromatic number exceeding $n^{0.016101}$, where $n = {m \choose
r}$ denotes the number of vertices in the graph. Further, for large
$k$, there exists a Kneser graph $K(m,r,t)$ that is $k$-vector
colorable but has chromatic number exceeding $n^{0.0717845}$.
\end{theorem}
\iflong
\begin{proof}
The basic idea is to improve the bound on the vector chromatic number
of the Kneser graph using an appropriately weighted version of the
characteristic vectors. We use weights $a$ and $-1$ to represent
presence and absence, respectively, of an element in the set
corresponding to a vertex in the Kneser graph, with appropriate
scaling to obtain a unit vector.  The value of $a$ that minimizes the
vector chromatic number can be found by differentiation and is
\[
A = -1 + \frac{mr}{r^2-rt} - \frac{mt}{r^2 -rt}
\]
Setting $a=A$ proves that the vector chromatic number is at most
\[
\frac{m(r-t)}{r^2 - mt}.
\]
At the same time, using Milner's Theorem proves that the exponent of
the chromatic number is at least
\[1
- - \frac{(m-t) \log \frac{2m}{m-t} + (m+t) \log \frac{2m}{m+t}} {2
\left( (m-r) \log \frac{m}{m-r} + r \log
\frac{m}{r} \right)}.
\]
By plotting these functions, we have shown that there is a set of
values with vector chromatic number 3 and chromatic number at least
$n^{0.016101}$. For large constant vector chromatic numbers, the
limiting value of the exponent of the chromatic number is roughly
$0.0717845$.
\end{proof}
\fi

\section{Conclusions}

The Lov\'asz number of a graph has been a subject of active study due to
the close connections between this parameter and the clique and
chromatic numbers. In particular, the following ``sandwich theorem'' was
proved by Lov\'asz~\cite{Lovasz:Shannon} (see Knuth~\cite{Knuth:sandwich}
for a survey).
\begin{equation}
\omega(G) \leq \vartheta(\overline{G}) \leq \chi(G).
\end{equation}
This led to the hope that the following question may have an affirmative
answer.
{\em Do there exist $\epsilon$, $\epsilon' > 0$ such that, for any graph $G$ on
$n$ vertices
\begin{equation}
\frac{\vartheta(\overline{G})}{n^{1 - \epsilon}} \leq \omega(G) \leq
\vartheta(\overline{G}) \leq \chi(G) \leq \vartheta(\overline{G}) \times
n^{1 - \epsilon'}?
\end{equation}
}
Our work in this paper proves a weak but non-trivial 
upper bound on the the chromatic number of $G$ in terms of
$\vartheta(\overline{G})$. However, this is far from achieving the bound
conjectured above and subsequent to our work, two results have
ended up answering this question negatively. Feige~\cite{feige} has
shown that for every $\epsilon>0$, there exist families of graphs for 
which $\chi(G) > \vartheta(\overline{G})n^{1 - \epsilon}$. Interestingly,
the families of graphs exhibited in Feige's work use
the construction of Section~\ref{sec:tightness} as a starting point.
Even more conclusively, the results of H{\aa}stad~\cite{hastad} and
Feige and Kilian~\cite{FK} have shown
that no {\em polynomial time computable function} approximates 
the clique number or chromatic number to within 
factors of $n^{1 - \epsilon}$, unless NP=RP.
Thus no simple modification of the $\vartheta$ function
is likely to provide a much better approximation guarantee.

In related results, Alon and Kahale
\cite{AK2} have also been able to use the semidefinite
programming technique in conjunction with our techniques to obtain algorithms
for computing bounds on the clique number of a graph with linear-sized
cliques, improving upon some results due to Boppana and
Halldorsson~\cite{boppana}.
Independent of our results, Szegedy~\cite{Szegedy:comm} has also shown
that a similar construction yields graphs with vector chromatic number
at most $3$ that are not colorable using $n^{0.05}$ colors. Notice
that the exponent obtained from his result is better than the one in
Section~\ref{sec:tightness}. Alon~\cite{alon:comm} has obtained a slight
improvement over Szegedy's bound by using an interesting variant of the
Kneser graph construction.
The main algorithm presented here has been derandomized in
a recent work of Mahajan and Ramesh~\cite{MR}.  By combining our
techniques with those of Blum~\cite{blum94}, Blum and
Karger~\cite{Karger:Coloring2} have given a 3-coloring algorithm with
approximation ratio $\Olog(n^{3/14})$.

\section*{Acknowledgments}
Thanks to David Williamson for giving us a preview of the MAX-CUT
result~\cite{GW:maxcut} during a visit to Stanford.  We are indebted 
to John Tukey and Jan Pedersen for their help in understanding 
multidimensional probability distributions.  Thanks to David 
Williamson and Eva Tardos for discussions of the duality theory 
of SDP. We thank Noga Alon, Don Coppersmith, Jon Kleinberg, 
Laci Lov\'asz and Mario Szegedy for useful discussions, and 
the anonymous referees for the careful comments.

%

\iflong\else
{\small
\fi

\iflong\else
}
\fi
\iflong
\comment{
\appendix

\section{An alternate analysis}
\label{appendix}

For a positive integer $k>1$, let $P_k(2,t)$ be the probability that two
unit vectors $a$ and $b$ in $\Re^2$ are separated by $t$ randomly chosen
vectors in $\Re^2$ given that the angle between $a$ and $b$ is $\theta$
such that $\cos \theta = -1/(k-1)$.
In this section we provide an alternate proof that $P_k =
\softo(t^{-k/(k-2)})$.

\begin{theorem}
$P_k(2,t) = \softo(t^{-k/(k-2)})$.
\end{theorem}

\begin{proof}
Without loss of generality, let $a$ be the vector $(\cos (\theta/2),
\sin (\theta/2))$ and let $b$ be the vector
$(\cos (\theta/2), -\sin (\theta/2))$. Let the $t$ randomly chosen vectors
be $\{(x_i, y_i)\}_{i=1}^t$ where the
$x_i$'s and the $y_i$'s are independent normal variables with mean $0$
and variance $1$.

Let $z_i$ be the projection of $(x_i,y_i)$ onto $a$, let $z'_i$ be
its projection onto $b$. Then $z_i$ and $z'_i$ are normal
variables with mean $0$ and variance $1$ given by
$$z_i = x_i \cos {\theta \over 2}
        + y_i \sin {\theta \over 2}$$
and
$$z'_i = x_i \cos {\theta \over 2}
        - y_i \sin {\theta \over 2}.$$

As usual we upper bound $P_k(2,t)$ by
$$t \cdot \Pr{z_1 \geq \max_i \{z_i\} \mbox{and} z'_1 \geq
\max_i \{z'_i\}},$$
which in turn is upper bounded by the quantity
\begin{equation}
\label{quantity}
\Pr{z_1 + z'_1 \geq \max_i \{z_i\} + \max_i \{z'_i\}}.
\end{equation}
We will pick a threshold $\tau$ carefully so that the following hold
\begin{equation}
\label{first}
\Pr{z_1 + z'_1 \geq 2\tau} = \softo(t^{-(2k-2)/(k-2)}),
\end{equation}
\begin{equation}
\label{second}
\Pr{\min\{\max_i \{z_i\}, \max_i \{z'_i\}\} <  \tau} \leq
2^{-\log^2 t + 1}.
\end{equation}
It is clear that using (\ref{first}) and (\ref{second})
the quantity in (\ref{quantity}) can be upper bounded by
$\softo(t^{-(2k-2)/(k-2)})$ which in turn suffices for the lemma.

The threshold $\tau$ is chosen to be
$$\sqrt{2 \ln t} - c \frac{\log\log t}{\sqrt{2 \ln t}},$$
where $c$ is a sufficiently large constant to be chosen
later. The inequality (\ref{first}) is obtained as follows:
\begin{eqnarray*}
\Pr{z_1 + z'_1 \geq 2\tau} & = & \Pr{2 \cos {\theta \over 2} x_1 \geq
2 \tau}\\
& = & \Pr{x_1 \geq {\tau \over \cos {\theta \over 2}}} \\
& \leq & \frac{\exp\left(\frac{-\tau^2}{2 \cos^2 {\theta \over 2}}
\right)}{\left({\tau \over \cos {\theta \over 2}}\right) (1 - o(1))} \\
& \leq & \softo\left(\exp\left(-\frac{\tau^2}{2 \cos^2 {\theta \over 2}}\right)\right) \\
& \leq & \softo\left(\exp\left(-\frac{\tau^2} {\cos \theta + 1}\right)\right) \\
& \leq & \softo\left(\exp\left(- \frac{2 \ln t - O(\log\log t)}{(k-2)/(k-1)} \right)\right)\\
& \leq & \softo(t^{-(2k-2)/(k-2)})
\end{eqnarray*}
The first of the above relationships is obtained by expressing $z_1$ and
$z'_1$ in terms of $x_1$ and $y_1$. The third line above is a standard
form on the tail probability of a normal distribution. The rest of the
calculation consists of elementary simplifications.

To obtain (\ref{second}) we use the fact that the
maximum of $t$ normally distributed variables has expected value
$$\sqrt{2 \ln t} - 1/2 {\log\log t \over {\sqrt{2 \ln t}}},$$
and the probability that it is $\alpha$ smaller than
its expectation is at most $\exp(-\exp(\alpha\sqrt{2 \ln t}))$ (see for instance~\cite{Aldous},
page 46).
Thus if we choose the constant $c$ in the definition of $\tau$ to be $5/2$,
then we obtain that
$$\Pr{\max_i \{z_i\} < \tau} \leq 2^{-\log^2 t}.$$
A similar bound can be obtained on $\Pr{\max_i \{z'_i\} < \tau}$.
\end{proof}
}
\fi
\end{document}